\documentclass[11pt]{article}
\usepackage{amssymb,amsmath,epsfig,xcolor,graphics,hyperref}
\DeclareMathOperator{\sech}{sech}
\def\one{{{{\rm 1} \kern -.19em {\rm l}}}}
\def\C{{{{\rm {\mbox{\small l}}} \kern -.50em {\rm C}}}}
\def\R{{{{\rm l} \kern -.15em {\rm R}}}}
\def\N{{{{\rm l} \kern -.15em {\rm N}}}}
\def\E{{{{\rm l} \kern -.15em {\rm E}}}}
\def\P{{{{\rm l} \kern -.15em {\rm P}}}} 
\def\Z{{{{\rm Z} \kern -.35em {\rm Z}}}}
\def\1{{{{\rm 1} \kern -.35em {\rm 1}}}}

\begin{document}
\begin{sloppypar}
\vspace*{0cm}
\begin{center}
{\setlength{\baselineskip}{1.0cm}{ {\Large{\bf 
Recursive Representation of Wronskians in Confluent Supersymmetric Quantum Mechanics 
 \\}} }}
\vspace*{1.0cm}
{\large{\sc{Alonso Contreras-Astorga}$^\dagger$} and {\sc{Axel Schulze-Halberg}$^\ddagger$}}
\indent \vspace{.3cm} 
\noindent \\	
Department of Mathematics and Actuarial Science and Department of Physics, Indiana
University Northwest, 3400 Broadway, Gary IN 46408, USA,\\ ${}^\dagger$E-mail:
aloncont@iun.edu 
\\ ${}^\ddagger$E-mail:
axgeschu@iun.edu, xbataxel@gmail.com 
\end{center}

\vspace*{.5cm}
\begin{abstract}
\noindent
A recursive form of arbitrary-order Wronskian associated with transformation functions in the 
confluent algorithm of supersymmetric quantum mechanics (SUSY) is constructed. With this recursive form regularity conditions for the generated potentials can be analyzed. Moreover, as byproducts we obtain new representations 
of solutions to Schr\"odinger equations that underwent a confluent SUSY-transformation.

\end{abstract} 

\section{Introduction}
The formalism of supersymmetry (SUSY) is one of the most significant methods for 
the construction of new solvable quantum models that feature a prescribed energy spectrum. Based 
on the mathematical concept of Darboux transformations that were first introduced in \cite{darboux}, 
the SUSY formalism interrelates quantum systems by means of linear differential operators (SUSY transformation). 
While such interrelated quantum systems are referred to as SUSY partners, the same terminology is commonly 
applied to their respective potentials. Since there is a vast amount of literature on the topic that encompasses many 
applications to particular quantum models, we refer the reader to the self-contained reviews 
\cite{cooper} \cite{djsusy} \cite{junker} and references therein. The SUSY formalism can be split into 
two different cases that we call standard and confluent SUSY algorithm. The standard algorithm is relatively well 
understood and used in most of the applications that can be found in the literature. Its application requires 
to determine solutions to the governing equation of the initial quantum system at pairwise different energies, these 
solutions are called transformation functions. The transformed system is then characterized by the Wronskian of the 
transformation functions. A comprehensive and very detailed review of the standard SUSY algorithm and applications 
can be found in \cite{bagrov}. In contrast to the latter standard algorithm, its confluent counterpart is less known. The simplest version of SUSY known as first-order SUSY has the restriction that only spectral modifications can be done below the ground state without introducing new singularities in constructed potential, in order to make more general manipulations of the spectrum an iteration can be done \cite{Baye87}. If the same transformation function is cleverly used during the iteration an energy level can be introduced or deleted, when only one iteration is considered the equivalence between this technique and the Levitan-Gelfand procedure can be proven \cite{Gelfand55,Moses80,Baye93}, this whole process is known as the confluent SUSY algorithm. This technique has been studied in particular contexts, for example the application of second and third-order 
SUSY transformations to certain quantum systems \cite{djinv} \cite{djenc} or the construction of orthogonal polynomials 
through arbitrary-order SUSY transformations \cite{grandati}, just to name a few. In the confluent SUSY algorithm, 
the transformation functions form a Jordan chain of the Hamiltonian that governs the initial 
quantum problem, that is, they are generalized eigenvectors of that Hamiltonian \cite{Andrianov04}. As such, the transformation 
functions admit both an integral and a differential representation \cite{xbatejp} \cite{bermudezthesis}, the 
relationship between which was reported on recently \cite{xbatalondiff} \cite{xbatint}. One of the principal open questions 
regarding the confluent SUSY algorithm concerns the regularity of the potential in the SUSY-transformed system. 
In order to not feature singularities, the Wronskian of the associated transformation functions is not allowed 
to have zeros inside the domain of the system. While for the standard SUSY algorithm this can be established by means of 
regularity conditions \cite{bagrov}, such conditions are only known for second and third-order confluent SUSY 
transformations \cite{djenc}, but not for higher orders. This is so because the regularity conditions emerge from the 
Wronskian of the transformation functions, the general form of which is known for second and third-order 
confluent SUSY transformations only. Therefore, the purpose of this work is to construct such a general form of 
the Wronskian for arbitrary-order confluent SUSY transformations. We approach this problem by 
constructing a recursive formula of the latter Wronskian for confluent SUSY transformations of arbitrary order. 
The advantages of this formula are twofold: first, we have a representation for the Wronskian which makes it 
much easier to derive regularity conditions for the SUSY-transformed potential. Second, since our formula allows to 
build a Wronskian from its lower-order counterparts, the calculation of transformation 
functions can be avoided. As a byproduct of our formula, we obtain alternative representations for SUSY-transformed 
solutions of the system's governing equations. In section 2 we summarize basic facts about the confluent SUSY algorithm. 
Section 3 is devoted to the construction of our recursive formula for the Wronskian associated with the 
transformation functions. In section 4 we present applications of our results in fourth and fifth-order confluent 
SUSY transformations.

\section{Supersymmetric quantum mechanics}
The SUSY formalism has its origin in the SUSY algebra, a graded Lie algebra of grade one. This algebra can be 
represented as the direct sum $S=P \oplus L$, where $P$ stands for the Poincare algebra and the space 
$L$ is spanned by two generators $Q_1$ and $Q_2$:
\begin{eqnarray}
Q_1 ~=~ \left(
\begin{array}{ccc}
0 & A^\dagger \\
A & 0
\end{array}
\right)
 \qquad \qquad \qquad
Q_2 ~=~ -i \left(
\begin{array}{ccc}
0 & A^\dagger \\
-A & 0
\end{array}
\right). \label{gen}
\end{eqnarray}
Here, $A$ and $A^\dagger$ are adjoint differential operators of order $n$ for two Schr\"odinger Hamiltonians $H_0$ and $H_1$, that is, 
\begin{eqnarray}
H_1~A^\dagger &=& A^\dagger ~H_0. \label{inter}
\end{eqnarray}
The generators $Q_1$, $Q_2$ satisfy the following anticommutator and commutator relations
\begin{eqnarray}
\left\{Q_i,Q_j \right\} ~=~ \delta_{ij}~H_S \qquad \qquad \qquad \left[Q_i,H_S \right] ~=~ 0,~~~i,j=1,2, \nonumber
\end{eqnarray}
where $H_S$ stands for the operator
\begin{eqnarray}
H_S &=& \left(
\begin{array}{ccc}
A^\dagger A & 0 \\
0 & A~A^\dagger
\end{array}
\right). \nonumber
\end{eqnarray}
We can relate this operator to the pair of Hamiltonians $H_0$ and $H_1$ in (\ref{inter}) 
by means of the operator factorization
\begin{eqnarray}
H_S &=& \prod_{j=0}^{n-1} \left[
\left(
\begin{array}{ccc}
H_1 & 0 \\ 0 & H_0
\end{array}
\right)-\lambda_j
\right], \nonumber
\end{eqnarray}
where $\lambda_j$, $j=0,...,n-1$, are complex-valued constants. Before we can comment further on the nature of these 
constants, we must distinguish the standard and the confluent SUSY algorithms. In the standard algorithm, the 
constants are pairwise different, that is, we have $\lambda_i \neq \lambda_j$ for $i,j=0,...,n-1$, $i \neq j$. A value 
$\lambda_j$ is associated with a solution $u_j$ of the Schr\"odinger equation $(H_0 -\lambda_j) u_j=0$. Observe that 
$\lambda_j$ is not required to be in the spectrum of $H_0$. In the confluent algorithm we have 
$\lambda_j = \lambda$, $j=0,...,n-1$, that is, all constants are equal to each other. The constant $\lambda$ is associated with 
$n$ solutions $u_j$, $j=0,...,n-1$, of the equation $(H_0 -\lambda)^n u_j=0$. If $\lambda$ belongs to the discrete 
spectrum of $H_0$, then the functions $u_j$ are called generalized eigenvectors or Jordan chain. Let us now 
assume that the operator $H_0$ admits a discrete spectrum $(E_j)$ and a family of associated eigenfunctions 
$(\Psi_j) \subset L^2(D)$, where $D$ is the domain of $V_0$. Then, for a fixed $j$ the function
\begin{eqnarray}
\Phi_j &=& A^\dagger \Psi_j, \label{darboux}
\end{eqnarray}
is an eigenfunction to the Hamiltonian $H_1$ for the spectral value $\lambda_j$, provided $A^\dagger \Psi_j \neq 0$. 
The operator $A^\dagger$ in 
(\ref{darboux}) can be expressed through Wronskians, as will be demonstrated in the subsequent two paragraphs. 
The Hamiltonian $H_1$ admits a complete set of eigenfunctions consisting of the $\Phi_j$ defined in (\ref{darboux}) and of 
solutions $\Phi_{\lambda_j}$ of the Schr\"odinger equation $(H_0 -\lambda_j) \Phi_{\lambda_j}=0$, provided 
$\lambda_j$ is an eigenvalue of $H_0$. In the two subsequent paragraphs we will specify computational details 
regarding the transformation (\ref{darboux}) and the potentials associated with our Hamiltonians $H_0$, $H_1$ for the 
standard and the confluent SUSY algorithm, respectively. Our starting point for reviewing the two SUSY algorithms is the 
one-dimensional stationary Schr\"odinger equation associated with the Hamiltonian $H_0$. We can write it in the form
\begin{eqnarray}
\Psi''+(E-V_0)~\Psi &=& 0, \label{schro}
\end{eqnarray}
where the energy $E$ is a real-valued constant and $V_0$ denotes the potential. 

\subsection{The standard SUSY algorithm} 
For a natural number $n$, assume that $u_0,u_1,...,u_{n-1}$ are solutions to equation (\ref{schro}), associated to the 
pairwise different energies $\lambda_0, \lambda_1,...,\lambda_{n-1}$, respectively. While the latter functions are 
often referred to as transformation functions or auxiliary solutions, their associated energies are usually called 
factorization energies. Now, the function
\begin{eqnarray}
\Phi &=& \frac{W_{u_0,u_1,...,u_{n-1},\Psi}}{W_{u_0,u_1,...,u_{n-1}}}, \label{susy2}
\end{eqnarray}
where each $W$ stands for the Wronskian of the functions in its index, is a solution to the transformed equation
\begin{eqnarray}
\Phi''+\left(E-V_n \right) \Phi &=& 0, \label{susy3}
\end{eqnarray}
the potential $V_n$ of which is related to its initial counterpart $V_0$ as
\begin{eqnarray}
V_n &=& V_0 -2~\frac{d^2}{dx^2}~ \log\left(W_{u_0,u_1,...,u_{n-1}} \right). \label{susy4}
\end{eqnarray}
Note that (\ref{susy2}) corresponds to (\ref{darboux}) and that $V_n$ is the potential associated with the 
Hamiltonian $H_1$. The function (\ref{susy2}) is called a SUSY transformation of order $n$ ($n$-SUSY transformation), also called 
Darboux or Darboux-Crum transformation. The latter two names refer to the mathematical origin of the 
SUSY algorithms, see \cite{darboux} and \cite{bagrov} for details. Note that the family 
$(u_1,u_2,...,u_n,\Psi)$ must be linearly independent in order to avoid that (\ref{susy2}) vanishes. If (\ref{schro}) is the 
governing equation of a spectral problem for the spectral parameter $E$, then a SUSY transformation of order $n$ can 
change the spectrum of the transformed problem by at most $n$ values, depending on the factorization energies
$\lambda_1, \lambda_2,...,\lambda_n$. Therefore, SUSY transformations are used for generating potentials that 
are associated with a prescribed spectrum (spectral design). While any $n$-SUSY transformation can be performed as 
$n$ successive first-order SUSY transformations, this increases the computational effort. 
Observe further that the transformation (\ref{susy2}) does not necessarily 
yield a physically meaningful result unless the transformation functions are properly chosen. Particularly in the case 
of transforming bound-state solutions, properties like normalizability and orthogonality of the transformation outcomes 
are not guaranteed.

\subsection{The confluent SUSY algorithm.}
In order to apply a 
$n$-th order confluent SUSY transformation to (\ref{schro}), 
we first determine $n+1$ functions 
$u_0,u_1,...,u_{n}$, that solve the following system of equations, 
\begin{eqnarray}
u_0''+(\lambda-V_0)~u_0 &=& 0 \label{j1} \\
u_j''+(\lambda-V_0)~u_j &=&- u_{j-1},~~~j=1,...,n, \label{j2}
\end{eqnarray}
introducing a real constant $\lambda$ that we assume to be different from $E$ in (\ref{schro}). Solutions to the system (\ref{j1}), (\ref{j2}) is commonly referred to as 
Jordan chain of order $n$. In the standard SUSY scheme, the auxiliary solutions must satisfy the initial Schr\"odinger 
equation at pairwise different energies \cite{bagrov}, which precisely constitutes the difference to the 
confluent algorithm that we are focusing on: here, all auxiliary solutions are associated with the same 
energy value $\lambda$. Now, once the system (\ref{j1}), (\ref{j2}) has been solved, we take a solution $\Psi$ of our initial 
Schr\"odinger equation (\ref{schro}) and construct the 
following functions $\Phi_n$ and $\chi_n$:
\begin{eqnarray}
\Phi_n ~=~ \frac{W_{u_0,...,u_{n-1},\Psi}}{W_{u_0,...,u_{n-1}}} \qquad \qquad \qquad 
\chi_n ~=~ \frac{W_{u_0,...,u_n}}{W_{u_0,...,u_{n-1}}}, \label{phi}
\end{eqnarray}
where the symbol $W$ stands for the Wronskian of the functions in its index. Then, 
$\Phi_n$ and $\chi_n$ are solutions to the following Schr\"odinger equations
\begin{eqnarray}
\Phi_n''+(E-V_n)~\Phi_n ~=~ 0 \qquad \qquad \qquad \chi_n''+(\lambda-V_n)~\chi_n ~=~ 0, \label{schrot}
\end{eqnarray}
recall that we required $\lambda \neq E$. The transformed potential $V_n$ is given by the expression
\begin{eqnarray}
V_n &=& V_0 - 2~\frac{d^2}{dx^2}~ \log\left(W_{u_0,u_1,...,u_{n-1}} \right). \label{pottrans}
\end{eqnarray}
As in case of the standard SUSY algorithm, confluent transformations deliver physically meaningful results only if the 
transformation functions are chosen appropriately. While the expressions for the transformed solutions (\ref{phi}) and its associated potential (\ref{pottrans}) 
look formally the same as in the conventional SUSY scheme, they are profoundly different 
due to the system (\ref{j1}), (\ref{j2}) that determines the transformation functions in the confluent case. These functions 
admit an integral and a differential representation. The first of these representations can be constructed by means of the 
variations-of-constants formula \cite{xbatejp}:
\begin{eqnarray}
u_j &=& \hat{u}-u_0~ \int\limits^x \left(\int\limits^t u_0~u_{j-1}~ds \right) \frac{1}{u_0^2}~dt,~~~j=1,...,n-1, 
\label{integral}
\end{eqnarray}
where $\hat{u}$ stands for any solution of the first equation (\ref{j1}). An alternative representation for the 
transformation functions involves parametric derivatives with respect to $E$ \cite{bermudezthesis}. Assuming that 
any solution of (\ref{j1}) is a function of the two variables $x$ and $\lambda$, we have
\begin{eqnarray}
u_j &=& \sum\limits_{k=0}^{j-1} \frac{\partial \hat{u}_k}{\partial \lambda^k} +\frac{1}{j!} \frac{\partial u_0}{\partial \lambda^j}, 
~~~j=1,...,n-1,
\label{u1rep2}
\end{eqnarray}
where $\hat{u}_k$, $k=0,...,n-2$, stand for arbitrary solutions of (\ref{j1}), including the trivial zero solution. Note that the 
representations (\ref{integral}) and (\ref{u1rep2}) are not equivalent, but related to each other \cite{xbatalondiff} 
\cite{xbatint}. Similar to the standard case, confluent SUSY transformations can be used for spectral design, but 
allow for the change of a single spectral value only. As such, the principal 
purpose of higher-order confluent SUSY transformations is the generation of new potentials that render the associated 
Schr\"odinger equation exactly-solvable. A typical application can be found in \cite{grandati}.

\section{Recursive representation of the Wronskian}
For the most part SUSY transformations are used to generate systems that are physically meaningful. As an 
important aspect of this, the transformed potential should either remain free of singularities or at least not receive 
additional singularities by undergoing the SUSY transformation. The principal quantity in the SUSY algorithms that 
controls the potential's singularities is the Wronskian of the transformation functions. As can be seen from 
(\ref{susy4}) and (\ref{pottrans}), any zero of the Wronskian contributes a singularity in the transformed potential. 
We are therefore interested in choosing the transformation function such that our Wronskian remains free of 
singularities. In case of second- and third-order confluent SUSY transformations this was achieved by means of 
constructing a closed-form expression of the Wronskian \cite{djenc}. In the following we will generalize this 
construction to arbitrary-order confluent SUSY transformations with the purpose to derive regularity conditions 
for the potentials resulting from such transformations.

\subsection{Construction of a recursion formula}
We will now derive a representation of our Wronskian through a particular case of the system (\ref{j1}), (\ref{j2}). 
More precisley, let us consider 
the following pair of equations
\begin{eqnarray}
\chi_{n-1}''+(\lambda-V_{n-1})~\chi_{n-1} &=& 0 \label{jn1} \\
\xi''+(\lambda-V_{n-1})~\xi &=&- \chi_{n-1}. \label{jn2}
\end{eqnarray}
We observe that these two equations are counterparts of (\ref{j1}), (\ref{j2}) for $n=1$. In contrast to the latter system, 
(\ref{jn1}) and (\ref{jn2}) apply to a Schr\"odinger equation that underwent a confluent SUSY transformation of order $n-1$. 
This Schr\"odinger equation is shown on the right side of (\ref{schrot}) if we replace $n$ by $n-1$. Let us now calculate 
the Wronskian of the functions $\chi_{n-1}$ and $\xi$. Its derivative reads
\begin{eqnarray}
W_{\chi_{n-1},\xi}' &=& \chi_{n-1}~\xi''-\chi_{n-1}''~\xi. \nonumber 
\end{eqnarray}
We replace the second derivatives by means of equations (\ref{jn1}), (\ref{jn2}). This gives
\begin{eqnarray}
W_{\chi_{n-1},\xi}' &=& \chi_{n-1} \left(V_{n-1}-\lambda\right) \xi-\chi_{n-1}^2-\left(V_{n-1}-\lambda\right) \chi_{n-1}~\xi \nonumber \\[1ex]
&=& -\chi_{n-1}^2. \nonumber
\end{eqnarray}
Consequently, integration on both sides reveals our Wronskian in the form
\begin{eqnarray}
W_{\chi_{n-1},\xi} &=& - \int\limits^x \chi_{n-1}^2~dt. \label{wron1}
\end{eqnarray}
This result is not surprising because it follows immediately from the known relation
\begin{eqnarray}
W_{u_0,u_1} &=& -\int\limits^x u_0^2~dt. \label{Wu0u1}
\end{eqnarray}
Let us now keep (\ref{wron1}) in mind, while we construct another form of the Wronskian 
$W_{\chi_{n-1},\xi}$. According to its definition, we have
\begin{eqnarray}
W_{\chi_{n-1},\xi} ~=~ \chi_{n-1}~\xi'-\chi_{n-1}'~\xi ~=~ \chi_{n-1} \left(\xi'
-\frac{\chi_{n-1}'}{\chi_{n-1}}~\xi\right).
\nonumber 
\end{eqnarray}
Now, the term in parenthesis can be interpreted as a SUSY transformation of first order, resulting in 
a solution $\chi_n$ of the Schr\"odinger equation on the right side of (\ref{schrot}). We have
\begin{eqnarray}
W_{\chi_{n-1},\xi} &=& \chi_{n-1}~\chi_n. \label{wron2}
\end{eqnarray}
After combining the results (\ref{wron1}) and (\ref{wron2}), we find
\begin{eqnarray}
\chi_{n-1}~\chi_n &=&  - \int\limits^x \chi_{n-1}^2~dt. \label{intpre}
\end{eqnarray}
Since the functions $\chi_n$ and $\chi_{n-1}$ obey the representation on the right side of (\ref{phi}), we can substitute 
the latter representation in (\ref{intpre}):
\begin{eqnarray}
\frac{W_{u_0,...,u_{n-1}}}{W_{u_0,...,u_{n-2}}}~\frac{W_{u_0,...,u_n}}{W_{u_0,...,u_{n-1}}} &=& 
-\int\limits^x \left(\frac{W_{u_0,...,u_{n-1}}}{W_{u_0,...,u_{n-2}}} \right)^2 dt. \nonumber
\end{eqnarray}
After minor simplifications we obtain the following result
\begin{eqnarray}
W_{u_0,...,u_n} &=& 
-W_{u_0,...,u_{n-2}}~\int\limits^x \left(\frac{W_{u_0,...,u_{n-1}}}{W_{u_0,...,u_{n-2}}} \right)^2 dt. \label{formula}
\end{eqnarray}
This is a recursive formula for the Wronskian of the transformation functions $u_j$, $j=0,...,n$, in the confluent 
SUSY algorithm. It is important to point out that the parameter $n$ in (\ref{formula}) can take arbitrarily high, but finite values. 
As an example, let us evaluate (\ref{formula}) for the particular case $n=2$. We obtain
\begin{eqnarray}
W_{u_0,u_1,u_2} &=& -u_0~\int\limits^x  \left(\frac{W_{u_0,u_1}}{u_0} \right)^2 dt. \nonumber
\end{eqnarray}
This coincides precisely with the result found in \cite{djenc}. Besides providing a representation of the 
Wronskian on its left side, our identity (\ref{formula}) can be used to determine SUSY-transformed potentials 
(\ref{pottrans}) without the need to calculate the transformation functions $u_j$, $j \geq 1$. We will illustrate 
this property in our application section 3.

\subsection{Factorization properties}
A closer inspection of our recursive formula (\ref{formula}) reveals that upon iterating it, the Wronskian on its left 
side takes a factorized form. We will now construct this form and determine its factors. As a first step, 
let us now use our formula (\ref{formula}) to 
replace the first factor $W_{u_0,...,u_n-2}$ on its right side. We obtain
\begin{eqnarray}
W_{u_0,...,u_n} &=& 
-\left[-
W_{u_0,...,u_{n-4}}~\int\limits^x \left(\frac{W_{u_0,...,u_{n-3}}}{W_{u_0,...,u_{n-4}}} \right)^2 dt
\right] \left[
\int\limits^x \left(\frac{W_{u_0,...,u_{n-1}}}{W_{u_0,...,u_{n-2}}} \right)^2 dt \right] \nonumber \\[1ex]
&=& W_{u_0,...,u_{n-4}} 
\left[ \int\limits^x \left(\frac{W_{u_0,...,u_{n-1}}}{W_{u_0,...,u_{n-2}}} \right)^2 dt \right]
\left[ \int\limits^x \left(\frac{W_{u_0,...,u_{n-3}}}{W_{u_0,...,u_{n-4}}} \right)^2 dt \right]. \nonumber
\end{eqnarray}
If this procedure is iterated, we obtain the following representation of our Wronskian
\begin{eqnarray}
W_{u_0,...,u_n} &=& 
\left\{
\begin{array}{lllll}
\displaystyle{(-1)^\frac{n}{2}~u_0~\prod\limits_{j=1 \atop j ~\mbox{\tiny{odd}}}^{n-1} 
\left[\int\limits^x \left(\frac{W_{u_0,...,u_{n-j}}}{W_{u_0,...,u_{n-j-1}}} \right)^2 dt \right]} & \mbox{if}~n~\mbox{is even}\\[1ex]
\displaystyle{(-1)^\frac{n-1}{2}~W_{u_0,u_1} \prod\limits_{j=1 \atop j ~\mbox{\tiny{odd}}}^{n-2} 
\left[\int\limits^x \left(\frac{W_{u_0,...,u_{n-j}}}{W_{u_0,...,u_{n-j-1}}} \right)^2 dt \right]}& \mbox{if}~n~\mbox{is odd}
\end{array}
\right\}. \nonumber
\end{eqnarray}
While this is already the sought factorized form, we can say a bit more about the actual factors. To this end, 
we observe that the quotient inside the integrals can be matched with the functions on the right side of 
(\ref{phi}), such that we have
\begin{eqnarray}
W_{u_0,...,u_n} &=& 
\left\{
\begin{array}{lllll}
\displaystyle{(-1)^\frac{n}{2}~u_0~\prod\limits_{j=1 \atop j ~\mbox{\tiny{odd}}}^{n-1} 
\left[\int\limits^x \chi_{n-j}^2~ dt \right]} & \mbox{if}~n~\mbox{is even}\\[1ex]
\displaystyle{(-1)^\frac{n-1}{2}~W_{u_0,u_1} \prod\limits_{j=1 \atop j ~\mbox{\tiny{odd}}}^{n-2} 
\left[\int\limits^x \chi_{n-j}^2~dt \right]}& \mbox{if}~n~\mbox{is odd}
\end{array}
\right\}. \label{fact}
\end{eqnarray}
Recall that according to (\ref{schrot}) and (\ref{pottrans}), a function $\chi_{n-j}$ solves the Schr\"odinger equation
\begin{eqnarray}
\chi_{n-j}''+\left[\lambda-V_0 + 2~\frac{d}{dx} \left(\frac{W'_{u_0,...,u_{n-j-1}}}{W_{u_0,...,u_{n-j-1}}}\right)-C\right] \chi_{n-j} ~=~ 0, \nonumber
\end{eqnarray}
where $j$ runs through odd natural numbers up to $n-1$ if $n$ is even or $n-2$ if $n$ is odd. In the final step we 
apply the recursive relation (\ref{intpre}) between the functions $\chi_j$, $j=1,...,n-1$ to (\ref{fact}), which is 
reduced to the simple form
\begin{eqnarray}
W_{u_0,...,u_n} &=& \prod\limits_{j=0}^{n} \chi_{j}, \label{fact2}
\end{eqnarray}
where we defined $\chi_0=u_0$. Since for a fixed $j$ the function $\chi_j$ can be written as a first-order 
SUSY transformation of $\chi_{j-1}$, the representation (\ref{fact2}) provides a factorization of our Wronskian 
in terms of SUSY transformation (or Darboux) operators. Such a factorization is known to exist for the 
standard SUSY formalism \cite{bagrov}. Note that this factorization as well as (\ref{fact2}), are not to be 
confused with the factorization due to Infeld and Hull \cite{infeld}, as the latter method does not apply to Wronskians, but 
to Hamiltonians.

\subsection{Representations of SUSY transformations}
In this section we will show that the recursive form of the Wronskian (\ref{formula}) leads to 
representations of SUSY transformations that can be used as an alternative to (\ref{phi}). 
The first of these representations is concerned with the solution $\chi_n$ of the Schr\"odinger equation on the 
right side of (\ref{schrot}). The latter equation admits its general solution to be written as a linear combination 
of two linearly independent functions, one of which is $\chi_n$ and the second function we call $\chi_n^\perp$. 
We can find $\chi_n^\perp$ through the reduction of order formula. This formula states that the solutions 
$\chi_n$ and $\chi_n^\perp$ are related as
\begin{eqnarray}
\chi_n &=& \chi_n^\perp~ \int\limits^x \frac{1}{\left(\chi_n^\perp\right)^2}~ dt,  \label{RO}
\end{eqnarray} 
their Wronskian being given by $W_{\chi_n^\perp, \chi_n}=1$. Keeping this in mind, we rewrite the function $\chi_n$ in 
(\ref{phi}) using our recursive formula (\ref{formula}). This gives
\begin{eqnarray}
\chi_n &=& \frac{W_{u_0, \dots ,u_{n-2}}}{W_{u_0, \dots ,u_{n-1}}} \int\limits^x \left( 
\frac{W_{u_0, \dots ,u_{n-1}}}{W_{u_0, \dots ,u_{n-2}}}  \right)^2 dt. \label{chi int}
\end{eqnarray} 
Now, comparison between \eqref{RO} and \eqref{chi int} leads to the conclusion that $\chi_n^\perp$ has the form
\begin{eqnarray}
\chi_n^\perp &=& \frac{W_{u_0,...,u_{n-2}}}{W_{u_0,...,u_{n-1}}}. \label{chi perp}
\end{eqnarray} 
Hence, the general solution of the Schr\"odinger equation on the right of (\ref{schrot}) can be expressed in terms 
of Wronskians. As a second application of our formula (\ref{formula}) we will now derive 
an alternative expression for the function $\Phi_n$ in \eqref{phi}. Before we start, let us point out that the definition of 
$\Phi_n$ in (\ref{phi}) depends on the Wronskian $W_{u_0, \dots,u_{n-1},\Psi}$. This will play a role below. 
Now, in order to obtain an alternative expression to \eqref{phi} for $\Phi_n$, a procedure similar to the one for obtaining 
\eqref{formula} can be followed. Let us consider the set of equations
\begin{eqnarray}
\chi_{n-1}'' +(\lambda-V_{n-1})~\chi_{n-1} &=& 0 \\[1ex]
\Phi_{n-1}'' + (E-V_{n-1})~ \Phi_{n-1} &=& 0.
\end{eqnarray}
The derivative of the Wronskian $W_{\chi_{n-1}, \Phi_{n-1}}$ obeys the form
\begin{eqnarray}
W_{\chi_{n-1}, \Phi_{n-1}}' ~=~ \chi_{n-1}~ \Phi_{n-1}'' - \chi_{n-1}'' ~\Phi_{n-1}~=~ (\lambda - E)~\chi_{n-1} ~\Phi_{n-1}.
\end{eqnarray}
After integrating both sides of the previous equation we obtain 
\begin{eqnarray}
W_{\chi_{n-1}, \Phi_{n-1}}&=& (\lambda - E)~\int\limits^x \chi_{n-1} \Phi_{n-1} dt. \label{WA}
\end{eqnarray}
Alternatively, using the definition of a Wronskian, we get
\begin{eqnarray}
W_{\chi_{n-1}, \Phi_{n-1}}~=~  \chi_{n-1}~ \Phi_{n-1}' - \chi_{n-1}'~ \Phi_{n-1} ~=~ 
\chi_{n-1} \left(\Phi_{n-1}'- \frac{\chi_{n-1}'}{\chi_{n-1}} ~\Phi_{n-1} \right)~ =~ \chi_{n-1} ~\Phi_{n} \label{WB}.
\end{eqnarray}
The left and right side of this identity can be combined to yield 
\begin{eqnarray}
\Phi_n &=& \frac{W_{\chi_{n-1}, \Phi_{n-1}}}{\chi_{n-1}}. \nonumber
\end{eqnarray}
In the final step we replace numerator and denominator on the right side by (\ref{WA}) and (\ref{phi}), 
respectively. This gives
\begin{eqnarray}
\Phi_n &=& (\lambda - E) ~\frac{W_{u_0, \dots, u_{n-2} }}{W_{u_0, \dots, u_{n-1}}}~ \int\limits^x 
\left( \frac{W_{u_0, \dots, u_{n-1}}}{W_{u_0, \dots, u_{n-2} }}  \right) 
\left( \frac{W_{u_0, \dots, u_{n-2},\Psi }}{W_{u_0, \dots, u_{n-2} }} \right) dt. \label{phinew}
\end{eqnarray}
Functions in \eqref{chi perp} and \eqref{phinew} are solutions of the Schr\"odinger equation \eqref{schrot}, and these 
representations are important when introducing integration constants in \eqref{formula} which will generate 
different SUSY partners potentials $V_n$. Observe that the solution index $n$ shown in (\ref{phinew}) can take arbitrarily 
high values, for example when enumerating an infinite set of bound state solutions. As such, there is ususally no 
limit of the sequence $(\Phi_n)$ in $L^2(D)$, where $D$ stands for the system's domain.

\section{Applications}
Let us now use our main results (\ref{formula}), (\ref{chi perp}) and (\ref{phinew}) in an example. 
We consider the Schr\"odinger equation
\begin{eqnarray}
\Psi''+\left( E + \frac{2}{\cosh^2(x)} \right) \Psi =0, \label{schroPT}
\end{eqnarray}
defined on the whole real line and equipped with boundary conditions $\lim\limits_{|x| \rightarrow \infty} \Psi(x)= 0$. 
We observe that (\ref{schroPT}) is a particular case of (\ref{schro}) if the potential $V_0$ is chosen as
\begin{eqnarray}
V_0(x)=-\frac{2}{\cosh^2(x)}. \label{V0PT}
\end{eqnarray}
This interaction is known as the P\"oschl-Teller potential \cite{flugge}. Even though there is already 
a vast amount of literature on this potential and its SUSY partners, it serves as a good toy model for our purposes, since 
the SUSY partners and their associated solutions of the Schr\"odinger equation are short enough to be stated in full form. 
The spectral problem governed by 
equation (\ref{schroPT}) and its boundary conditions admits a discrete spectrum that contains the single value 
$E=-1$. This eigenvalue is associated with a bound state-solution $\Psi \in L^2 \left(\mathbb{R} \right)$, given by
\begin{eqnarray}
\Psi = \frac{1}{\sqrt{2} \cosh(x)}.  \label{psi PT}
\end{eqnarray}
Note that the numerical factor is included to ensure correct normalization. In order to perform higher-order SUSY transformations we need the transformation functions $u_0$ and 
$u_j$, $j=1,2,...$, that are solutions of \eqref{j1} and \eqref{j2}, respectively. Once we have selected the 
function $u_0$ we can construct the functions $u_j$, $j=1,2,...$, through their integral representation \eqref{integral}. 
For the present example we use 
\begin{eqnarray}
u_0= \sqrt{2 \kappa}~ \exp(\kappa x)\left[\tanh(x)-\kappa \right], \label{u0PT}
\end{eqnarray}
where $\kappa = \sqrt{-\lambda}$, recall that $\lambda$ is the factorization energy from (\ref{j1}), (\ref{j2}). Throughout this 
example we will assume that $\lambda<0$, such that $\kappa>0$. Notice that the solution $u_0$ of \eqref{j1} vanishes for $x \rightarrow -\infty$ if $\kappa>0$, 
this will be important for the regularity of the new potentials generated by our SUSY transformation. 
To perform fourth and fifth-order transformations the needed remaining transformation functions $u_1,u_2,u_3$, 
obtained using \eqref{integral}, are:
\begin{eqnarray}
u_1 & = &  -\frac{\exp(\kappa x)}{\sqrt{2 \kappa^3}}  \left[\kappa ^2 x- (\kappa  x -1) \tanh (x)\right]  \label{u1PT}  \\
u_2 & = &  \frac{\exp(\kappa  x)}{4 \sqrt{2 \kappa^7}}   \left\{\kappa ^2 x (1-\kappa  x)+[ \kappa  x (\kappa  x-3)+3] \tanh (x)\right\} \label{u2PT} \\
u_3 & = & \frac{\exp(\kappa  x)}{24 \sqrt{2 \kappa ^{11}}}  \left\{ -\kappa ^2 x \left[ \kappa  x (\kappa  x-3)+3\right] + 
\left[ \kappa  x (\kappa^2  x^2 -6 \kappa x +15 ) -15 \right]  \tanh (x)\right\} \label{u3PT}
\end{eqnarray}
Observe that in principle we need another transformation function for the fifth-order case. However, we will 
show that using our representations (\ref{formula}) and (\ref{phinew}), we can avoid the computation of $u_4$. 
Let us point out that in principle we do not need to determine any transformation function except for $u_0$. Instead, we 
can iterate our formula (\ref{formula}) to successively generate the Wronskians $W_{u_0,u_1}$, $W_{u_0,u_1,u_2}$ 
and $W_{u_0,u_1,u_2,u_3}$. Since these three steps involve a large amount of calculations, we restrict ourselves here to 
showing the last step only in the next section.

\subsection{Fourth-order SUSY transformation}
\begin{figure}[t]
\begin{center}
\includegraphics[width=7 cm]{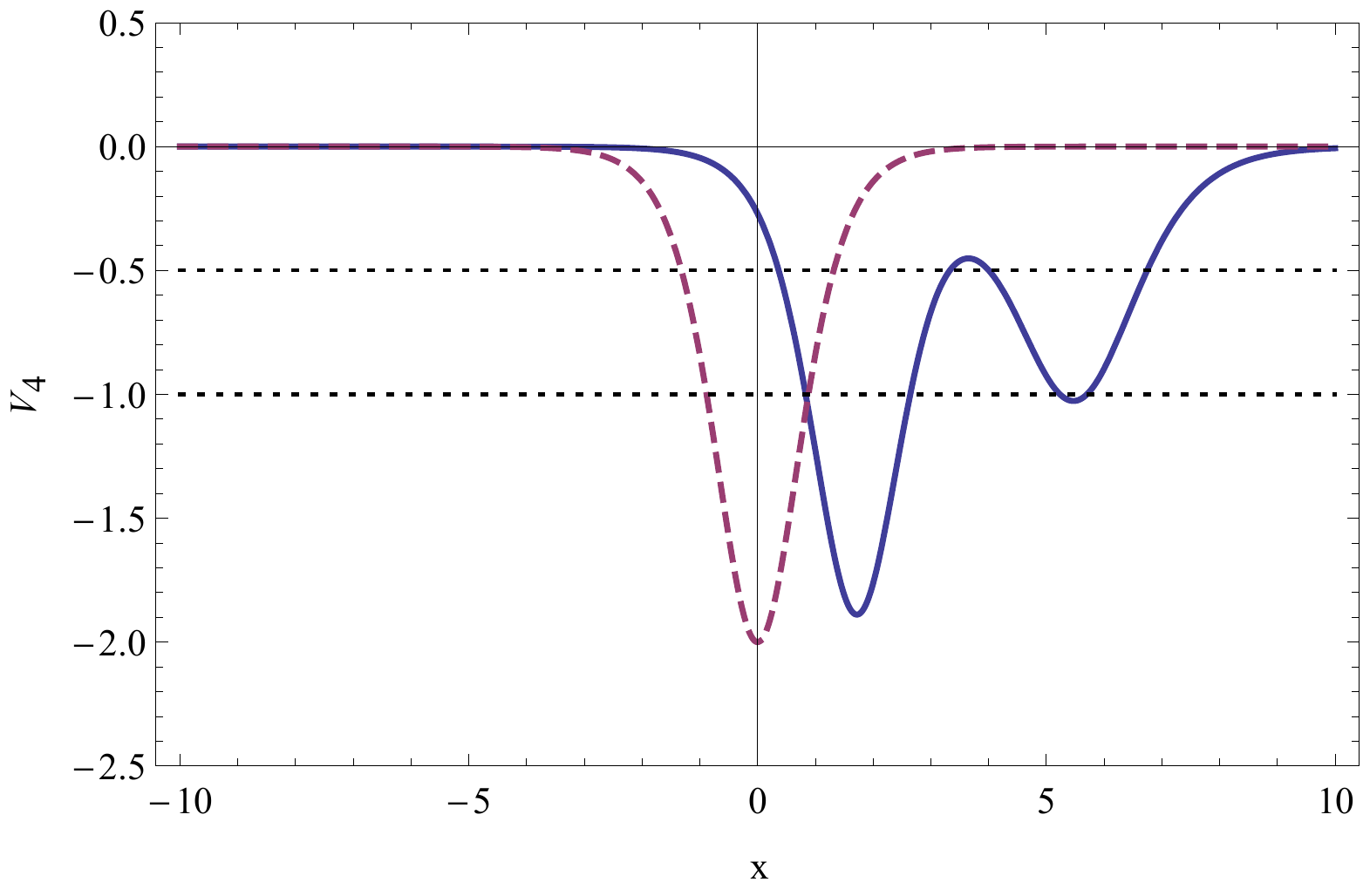} \hspace{.4cm}
\includegraphics[width=7 cm]{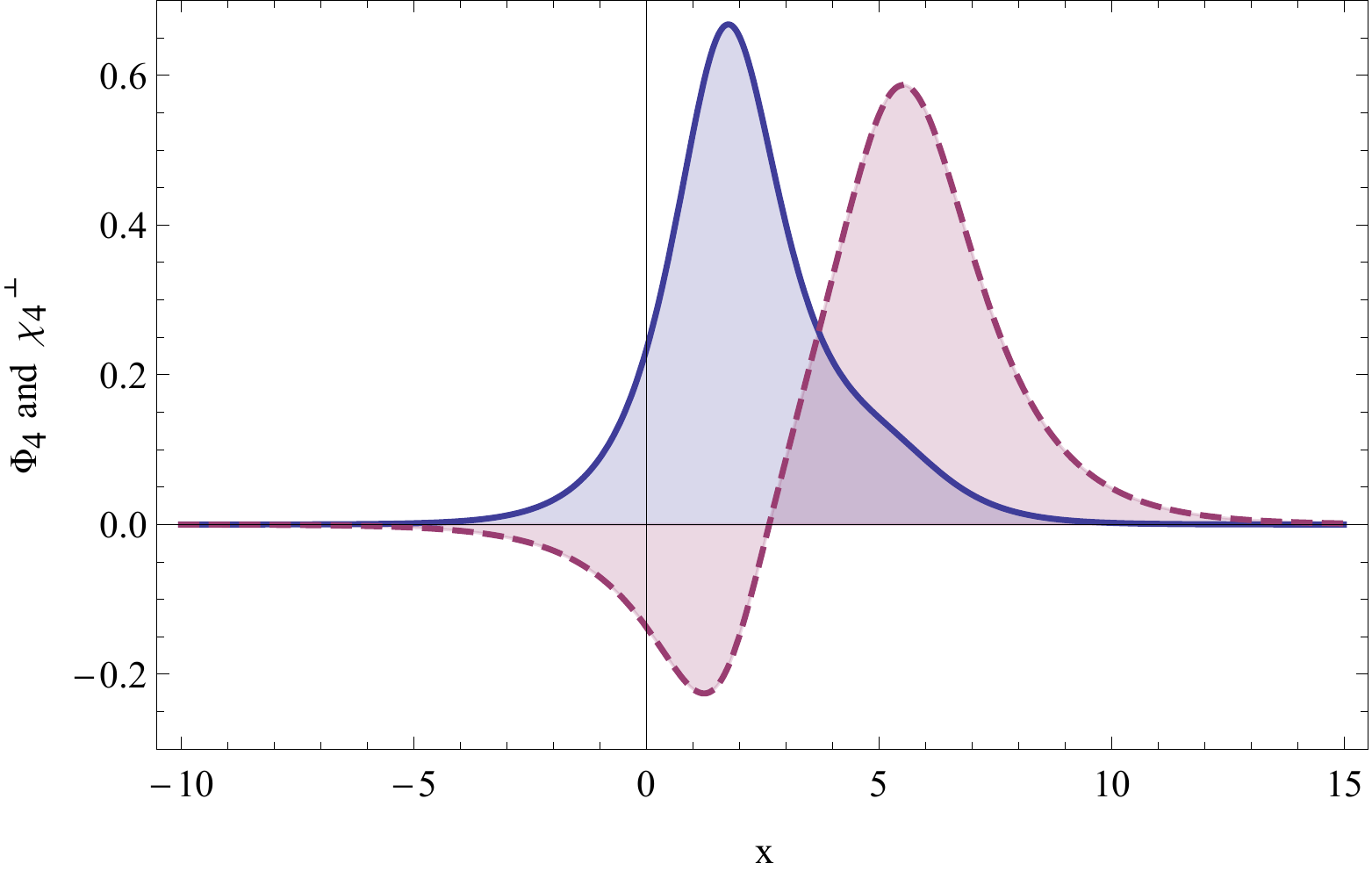}
\caption{Left: Fourth-order SUSY partner of the hyperbolic P\"oschl-Teller potential (blue solid curve). 
The P\"oschl-Teller potential is plotted as reference (purple dashed curve), the parameters used are 
$C_a = 50,~ \kappa = 1/\sqrt{2}$. Right: Eigenfunctions $\Phi_4$ and $\chi_4^\perp$ corresponding to the potential on the left. 
}
\label{PT potentials}
\end{center}
\end{figure}

To perform a fourth-order SUSY transformation to our equation (\ref{schroPT}), we will use the functions 
$u_0$ and $u_1, ~u_2,~u_3$ that are given in \eqref{u0PT} and (\ref{u1PT})-(\ref{u3PT}), respectively. Before applying 
the actual transformation, we need to make sure that the resulting transformed potential $V_1$ in (\ref{schrot}) is free 
of singularities on the real line. Since this is guaranteed if the Wronskian of the transformation functions does not 
vanish, we must analyze this Wronskian. Its explicit form can be obtained in two different ways, either by direct 
calculation using  \eqref{u0PT}-(\ref{u3PT}) or through our recursive formula \eqref{formula}. Using the latter 
formula, we obtain
\begin{eqnarray}
W_{u_0,u_1,u_2,u_3}=-W_{u_0,u_1}\cdot \int^x \left( \frac{W_{u_0,u_1,u_2}}{W_{u_0,u_1}} \right)^2 dt. 
\label{PTWronskian a}
\end{eqnarray}
Observe that the right side of this identity does not contain the transformation function $u_3$.  Since the integration 
is indefinite, it involves an arbitrary additive constant that we call $C_a$. We can therefore rewrite 
(\ref{PTWronskian a}) in a somewhat intuitive matter as
\begin{eqnarray}
W_{u_0,u_1,u_2,u_3} &= & -W_{u_0,u_1}\cdot  \left[ C_a + \int\limits^x \left( \frac{W_{u_0,u_1,u_2}}{W_{u_0,u_1}} \right)^2 dt \right]
 \label{PTWronskian b}\\[1ex]
& =& \exp(2\kappa x)\left[1+\kappa ^2-2 \kappa  \tanh (x)\right] \nonumber \\[1ex]
& \times &  \left[ C_a+\exp(2 \kappa x) \frac{ \left(\kappa ^4+6 \kappa ^2+1\right) \cosh (x)-4 \left(\kappa ^3+
\kappa \right) \sinh (x)}{16 \kappa ^4 \left[\left(\kappa ^2+1\right) \cosh(x)-2 \kappa  \sinh (x) \right]}\right]. 
\label{PTWronskian x}
\end{eqnarray}
Since in this relatively simple example we were able to calculate the transformation functions (\ref{u0PT})-(\ref{u3PT}), 
let us verify the result (\ref{PTWronskian x}) by calculating the Wronskian $W_{u_0,u_1,u_2,u_3}$ directly. We obtain
\begin{eqnarray}
W_{u_0,u_1,u_2,u_3} &=& \frac{1}{16 \kappa^4}~\exp(4\kappa x) \left[1+6\kappa^2+\kappa^4-4~(k+k^3)~\tanh(x)
\right]. \label{wronex}
\end{eqnarray}
One verifies by direct calculation that this is a special case of (\ref{PTWronskian x}), obtained by substituting $C_a=0$ and 
simplifying. It is clear that our result (\ref{wronex}) cannot contain any free constants because all transformation 
functions are determined. Since in (\ref{PTWronskian x}) we did not use our function $u_3$, the resulting expression for 
the Wronskian is determined up to an integration constant, which allows adjustment to yield a regular potential. 
Recall that we need this Wronskian to be free of zeros 
in order to avoid singularities in the SUSY-transformed potential. A further analysis shows that the integral in 
\eqref{PTWronskian b} is a monotonically increasing function that tends to zero when $x$ goes to $- \infty$. 
As a direct consequence, if the condition 
$C_a>0$ is fulfilled, then the term in brackets on the right side of (\ref{PTWronskian x}) does not have any zeros. 
The remaining factor $W_{u_0,u_1}$ in the same expression can be shown to not contribute any zeros either, which 
is a direct consequence of $u_0$ vanishing for $x \rightarrow -\infty$. We can now construct SUSY partner potentials $V_4$ 
using \eqref{pottrans} with $n=4$, $V_0$ as in \eqref{V0PT} and $W_{u_0,u_1,u_2,u_3}$ as 
in \eqref{PTWronskian b}, note that every nonnegative value of the integration constant $C_a$ will lead to a 
different regular SUSY-transformed potential. The explicit expression of the family of partner potentials $V_4$ is 
not displayed here 
because of its length, but since all functions involved are shown it can be seen that the final 
expression will involve only exponential and hyperbolic functions. A particular case of a potential 
$V_4$ can be found in figure \ref{PT potentials}. Next we determine the SUSY-transformed counterpart of 
(\ref{psi PT}) by means of (\ref{phi}) for $n=4$ by substituting (\ref{psi PT}) and (\ref{PTWronskian x}) into the 
representation (\ref{phinew}). 
We obtain an eigenfunction $\Phi_4 \in L^2\left(\mathbb{R} \right)$, associated with the eigenvalue $E=-1$, the 
explicit form of which reads
\begin{eqnarray}
\Phi_4 = \frac{\left(\kappa ^2-1\right)^2 \sech(x) \left[16 C_a \kappa ^4-\left(\kappa ^2-1\right) \exp (2 \kappa  x)\right]}{\sqrt{2} \left\{16 C_a \kappa ^4 \left[ \kappa ^2-2 \kappa  \tanh (x)+1\right]+\exp(2 \kappa  x) \left[\kappa ^4+6 \kappa ^2-4 \left(\kappa ^3+\kappa \right) \tanh (x)+1\right]\right\}}.
\end{eqnarray}
The second eigenfunction of this 2-level system is $\chi_4^\perp$ given by \eqref{chi perp} with $\lambda$ as corresponding eigenvalue and it is given by 
\begin{eqnarray}
\chi_4^\perp = \frac{\exp(\kappa  x) \left[\kappa  \left(\kappa ^2+3\right) -\left(3 \kappa ^2+1\right) \tanh (x)\right]}{2 \sqrt{2 \kappa^3}  \left[\kappa ^2-2 \kappa  \tanh (x)+1\right] \left\{C_a+\frac{\exp (2 \kappa  x) \left[  \left(\kappa ^4+6 \kappa ^2+1\right) \cosh (x)-4 \left(\kappa ^3+\kappa \right) \sinh (x)\right]}{16 \kappa ^4 \left[\left(\kappa ^2+1\right) \cosh (x)-2 \kappa  \sinh (x)\right]}\right\}}.
\end{eqnarray}
In contrast to the initial problem that admits a single eigenvalue, the discrete spectrum of the transformed problem 
associated with the potential $V_4$, contains two eigenvalues. More precisely, the discrete spectrum is given by 
$\{-1, \lambda \}$, i. e., the SUSY transformation added the 
eigenvalue $\lambda$ to the discrete spectrum. It is worth noticing that for any even transformation 
$\lambda$ can be greater or less than the ground state energy $E=-1$.  
The left part of figure \ref{PT potentials} shows a SUSY partner $V_4$ (blue solid curve) with the parameters 
$C_a = 50,~ \kappa = 1/\sqrt{2}$ and the P\"oschl-Teller potential $V_0$ (see \eqref{V0PT}) plotted as a reference 
(purple dashed curve), horizontal lines corresponding to $E=-1$ and $\lambda = -1/2$ are also plotted (black dotted curves). 
On the right of the figure, the only two eigenfunctions (normalized) of the corresponding Schr\"odinger equation for $V_4$ are plotted, $\Phi_4$ with eigenvalue $E=-1$ (blue continuous curve) and the first excited state $\chi_n^\perp$ (purple dashed curve) with $\lambda = -1/2$ as corresponding eigenvalue. 

\subsection{Fifth-order SUSY transformation}
\begin{figure}[t]
\begin{center}
\includegraphics[width=7 cm]{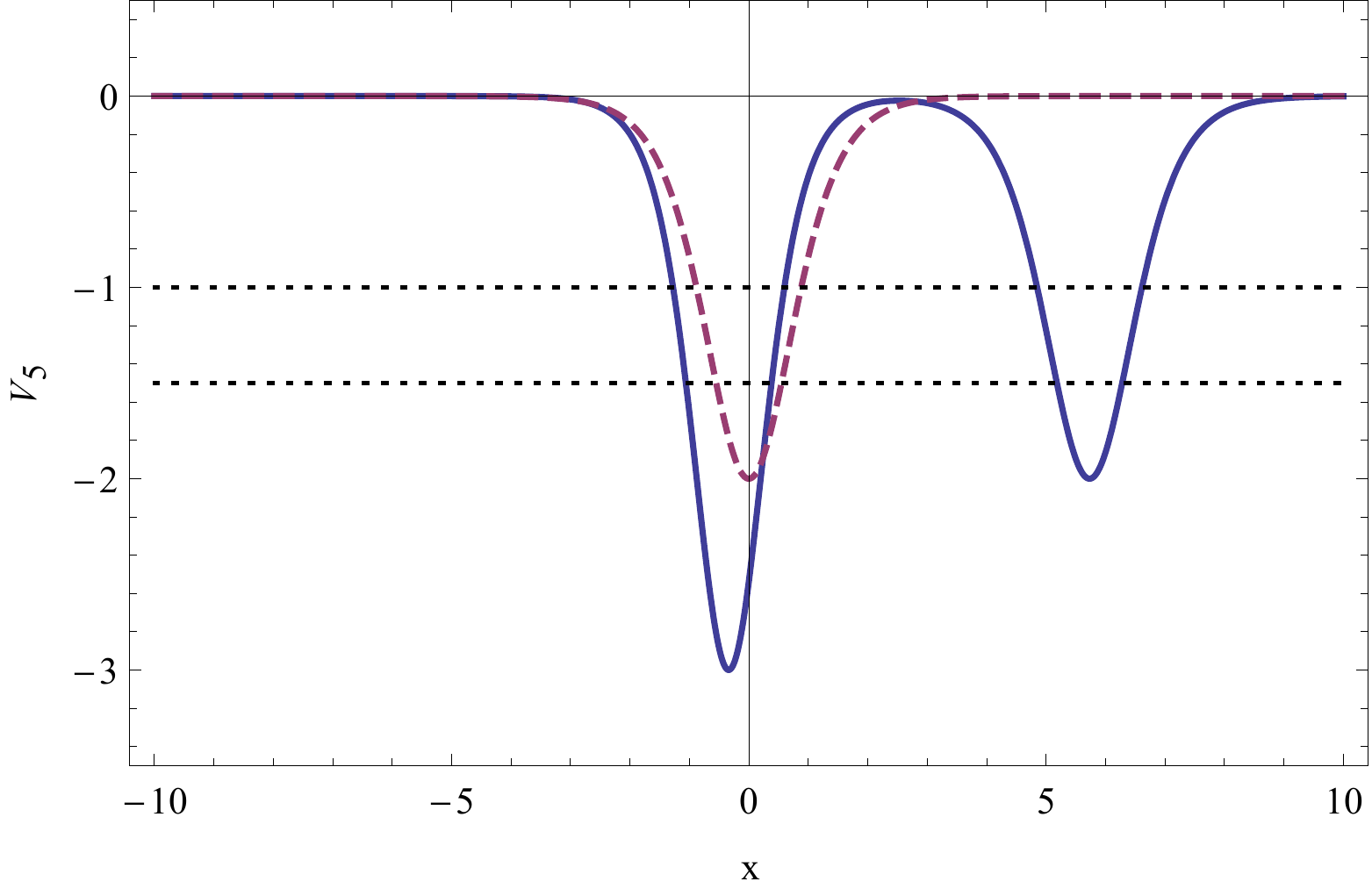} \hspace{.4cm}
\includegraphics[width=7 cm]{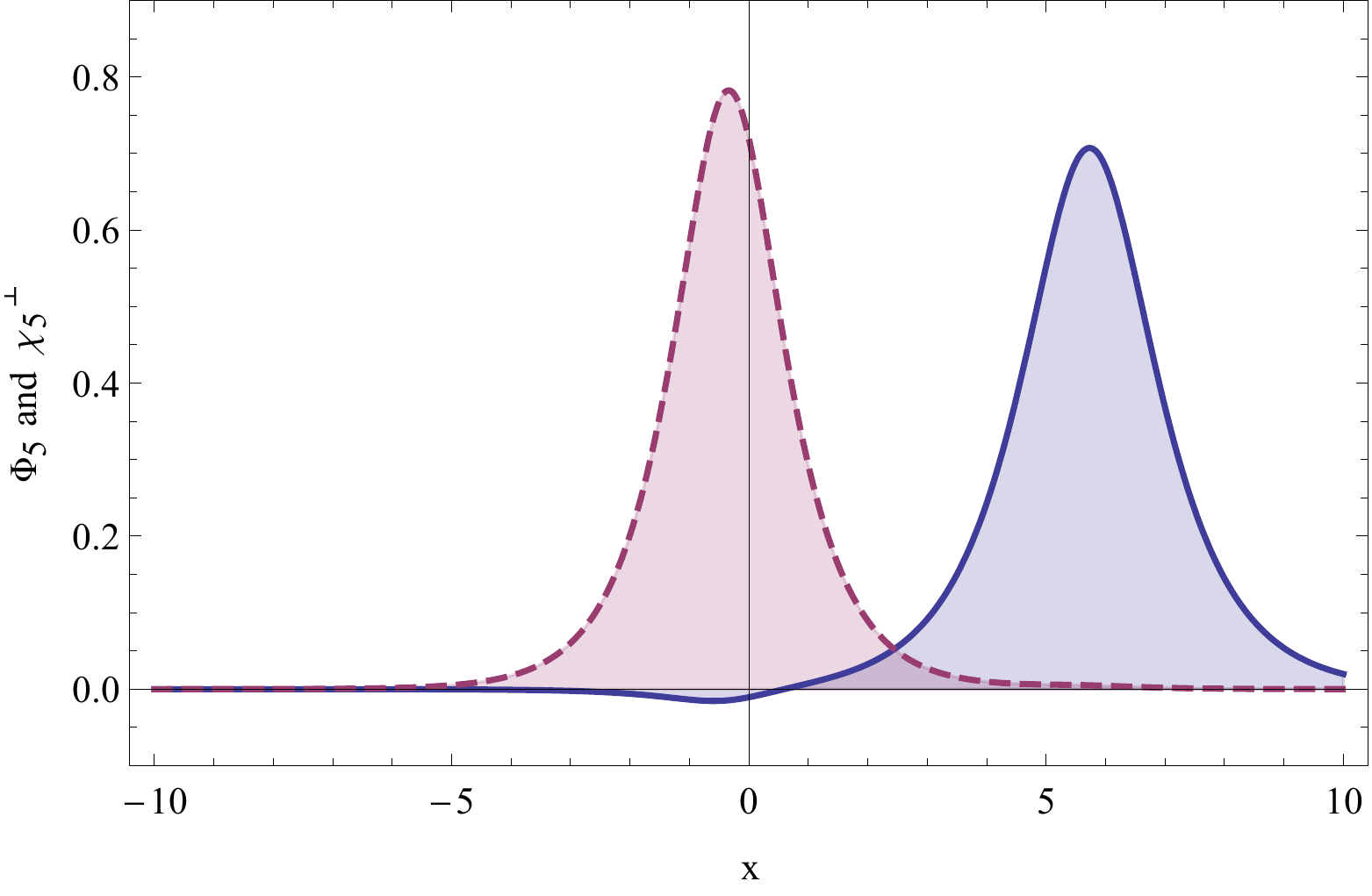}
\caption{Left: Fifth-order SUSY partner of the hyperbolic P\"oschl-Teller potential (blue solid curve). The P\"oschl-Teller 
potential is plotted as reference (purple dashed curve). The parameters used are $C_b = 0.01,~ \kappa = \sqrt{3/2}$. 
Right: functions $\Phi_5$ (blue solid curve) and $\chi_5^\perp$ (purple dashed curve) corresponding to the potential on the left. 
}
\label{PT 5 potentials}
\end{center}
\end{figure}

For the construction of a SUSY partner potential to (\ref{V0PT}) using a fifth-order transformation, we proceed as in the previous 
example. As a first step, we compute the Wronskian $W_{u_0, u_1, u_2, u_3, u_4}$ of the transformation functions and 
derive a condition for it to be free of zeros. We cannot find this Wronskian directly because we do not have the 
transformation function $u_4$. Instead, we will use once more our recursive representation (\ref{formula}) that only 
requires knowledge of the transformation functions $u_0,u_1,u_2$ and $u_3$. Upon 
substituting (\ref{u0PT})-(\ref{u3PT}) and including a constant of integration $C_b$, we arrive at 
\begin{eqnarray}
W_{u_0, u_1, u_2, u_3, u_4}&=& - W_{u_0,u_1,u_2} \cdot \left[C_b+ \int\limits^x 
\left( \frac{W_{u_0,u_1,u_2,u_3}}{W_{u_0,u_1,u_2}} \right)^2 dt \right]\label{Wu4pre} \\[1ex]
&\hspace{-1.7cm} = & \hspace{-1cm} \frac{\exp(3 \kappa x)}{2 \sqrt{2 \kappa^3}} \left[-\kappa  \left(\kappa ^2+3\right)+(3 \kappa^2+1)\tanh(x)  \right] \nonumber 
\\[1ex]
& \hspace{-1.7cm} \times &  \hspace{-1cm} \left\{C_b + \exp(2 \kappa x) \frac{\kappa  \left(\kappa ^4+10 \kappa ^2+5\right) \cosh (x)-\left(5 \kappa ^4+10 \kappa ^2+1\right) \sinh (x)}{64 \kappa ^6 \left[ \kappa  \left(\kappa ^2+3\right) \cosh (x)-\left(3 \kappa ^2+1\right) \sinh (x)\right]}   \right\}
. \label{Wu4}
\end{eqnarray} 
Using an argument similar to the fourth-order case, we can ensure that this Wronskian will not 
vanish on the real axis by imposing the constraint $C_b \geq 0$. Furthermore, when the Wronskian 
$W_{u_0,u_1,u_2}$ on the right side of (\ref{Wu4pre}) 
is factored using again \eqref{formula}, it can be observed that $u_0$ is a factor, 
thus $u_0$ has to be chosen without any zeros. According to the Sturm oscillatory theorem \cite{Berezin} 
a necessary condition to have a non vanishing function $u_0$ is that $\lambda$ must be less than or 
equal to the ground state energy $E=-1$. Using this setting, a fifth-order SUSY potential $V_5$ that is free of singularities 
can then be generated using \eqref{pottrans}, \eqref{V0PT} and \eqref{Wu4}, 
note that an explicit expression of $V_5$ is not displayed due to its length. A particular case of $V_5$ is shown in 
figure \ref{PT 5 potentials}. Let us point out that for the construction of this potential we did not have to compute 
the transformation function $u_4$ by using the recursive formula (\ref{formula}). It remains to determine the solutions 
associated with the transformed potential $V_5$. To this end, we first observe that due to the constraint 
$\lambda \leq -1$, guaranteeing regularity of $V_5$, the solution associated with the lowest eigenvalue is 
given by $\chi_5^\perp$. This function can be constructed using  \eqref{chi int} with  
$W_{u_0, u_1, u_2, u_3, u_4}$ as in \eqref{Wu4}, its explicit form is then
\begin{eqnarray}
\chi_5^\perp &=& -\left\{8 \sqrt{2 \kappa ^7} \exp(\kappa  x) \left[\left(\kappa ^4+6 \kappa ^2+1\right) \cosh (x)-4 \kappa  
\left(\kappa ^2+1\right) \sinh (x)\right] \right\} \times \nonumber \\[1ex]
&\times& \left\{64 C_b \kappa ^6 \left[\kappa  \left(\kappa ^2+3\right) \cosh (x)-\left(3 \kappa ^2+1\right) 
\sinh (x)\right]+\exp(2 \kappa  x)  \times \right. \nonumber \\[1ex]
&\times& \left.
\left[  \left(\kappa ^5+10 \kappa ^3+5\kappa \right) \cosh (x)-
\left(5 \kappa ^4+10 \kappa^2+1\right) \sinh (x)\right] \right\}^{-1}. \nonumber
\end{eqnarray}
The remaining solution $\Phi_5$, representing the first excited state of the system, can be obtained by 
evaluating \eqref{phinew}. After some elementary simplification we arrive at
{\footnotesize \begin{eqnarray}
\Phi_5 = \frac{\left(\kappa ^2-1\right)^3 \text{sech}(x) \left[\left(\kappa ^2-1\right) 
\exp (2 \kappa  x)-64 C_b \kappa ^6\right]}{\sqrt{2} \left\{ 64 C_b \kappa ^6 \left[ \kappa ^3+3\kappa -\left(3 \kappa ^2+1\right) 
\tanh (x)\right]+\exp(2 \kappa  x) \left[\kappa ^5+10 \kappa ^3+5 \kappa -\left(5 \kappa ^4+10  \kappa ^2+1\right) 
\tanh (x)\right]\right\}}. \nonumber 
\end{eqnarray} 
}
The discrete spectrum of the transformed problem for the potential $V_5$ is $\{\lambda, -1 \}$, that is, it contains two 
eigenvalues, one of which was created through the SUSY transformation. In the left part of figure \ref{PT 5 potentials} we see a fifth-order SUSY partner $V_5$ (blue solid curve), 
generated with the parameters $C_b = 0.01$ and $\lambda = -3/2$, the P\"oschl-Teller potential is also 
displayed (purple dashed curve) and horizontal lines for the values $E=-1$ and $\lambda = -3/2$ are also plotted 
(black dotted lines). On the right of the same figure, the eigenfunctions $\Phi_5$ (blue solid curve) 
and $\chi_5^\perp$ (purple dashed curve) corresponding the eigenvalues $E=-1$ and $\lambda = -3/2$, respectively, 
are shown. This system is governed by a non-symmetric double well potential, where a particle in the ground state is localized 
around the deeper well meanwhile a particle in the excited level is localized in the second well with a low probability to be 
found inside the deeper well.  

\section{Concluding remarks}
We have constructed a recursive formula for the Wronskian of transformation functions, as they appear in the confluent 
SUSY algorithm. As byproducts, we obtained factorizations of the Wronskian and 
alternative representations of SUSY transformations. The main application of our results lies in the facilitation of 
establishing regularity conditions for potentials that were obtained through confluent SUSY transformations. 
While a complete analysis of such regularity conditions is beyond the scope of the present work, our results 
can prove useful when dealing with particular quantum systems. Besides the already well-known P\"oschl-Teller model 
studied in section 4, our Wronskian representation (\ref{formula}) and the emerging regularity conditions are applicable 
in different contexts. As a recent interesting example for such an application, let us mention the work \cite{fernandez}. 
Here, a fourth-order confluent SUSY algorithm is applied to the free-particle system, generating 
Neumann-Wigner-type potentials that allow for the existence of so-called bound-states in the continuum. Since the 
underlying confluent  SUSY transformation is of order four, the Wronskian of the transformation functions can be 
represented and analyzed using our recursive formula (\ref{formula}).

\end{sloppypar}
\end{document}